\documentclass[a4paper,11pt]{article}
\usepackage{pos}

\title{Topological studies of light-flavor hadron production in high multiplicity pp collisions with ALICE at the LHC}
\ShortTitle{Topological studies of light-flavor hadron production with ALICE}

\author*{Sushanta Tripathy}
\author{ALICE collaboration}

\affiliation{Instituto de Ciencias Nucleares, Universidad Nacional Aut\'onoma de M\'exico,\\
  Apartado Postal 70-543, Ciudad de M\'exico 04510, M\'exico}

\emailAdd{sushanta.tripathy@cern.ch}

\abstract{Recent measurements in high-multiplicity pp and p-A collisions have revealed that these small collision systems exhibit collective-like behaviour, formerly thought to be achievable only in heavy-ion collisions. To understand the origins of these unexpected phenomena, event shape observables can be exploited, as they serve as a powerful tool to disentangle soft and hard contributions to particle production. Here, results on the production of light flavor hadrons for different classes of unweighted transverse spherocity ($S_{\rm 0}^{p_{\rm T}= 1}$) and relative transverse activity ($R_{\rm{T}}$) in high multiplicity pp collisions at $\sqrt{s}$ = 13~$\textrm{TeV}$ measured with the ALICE detector are presented. Hadron-to-pion ratios in different $S_{\rm 0}^{p_{\rm T}= 1}$ and $R_{\rm{T}}$ classes are also presented and compared with state-of-the-art QCD-inspired Monte Carlo event generators. The evolution of charged particle average transverse momentum ($\langle p_{\rm T}\rangle$) with multiplicity and $S_{\rm 0}^{p_{\rm T}= 1}$ is also discussed. In addition, the system size dependence of charged particle production in pp, p--Pb, and Pb--Pb collisions at $\sqrt{s_{\rm NN}}$= 5.02 TeV is presented. Finally, within the same approach, we present a search for jet quenching behavior in small collision systems.}

\FullConference{%
  40th International Conference on High Energy physics - ICHEP2020\\
  July 28 - August 6, 2020\\
  Prague, Czech Republic (virtual meeting)
}

\begin{document}
\maketitle
\section{Introduction}
Recent ALICE~\cite{ALICE:2017jyt} measurements show a smooth evolution of strange to non-strange particle ratios across different colliding systems (pp, p--Pb, and Pb--Pb) as a function of charged-particle multiplicities, which may point towards a common underlying physics mechanism across collision systems. Collective-like effects have also been observed in small collision systems, but no jet quenching signatures have been reported in pp or p--Pb collisions yet~\cite{Nagle:2018nvi}. These observed behaviors are quite challenging for currently popular event generators, like PYTHIA8~\cite{Sjostrand:2014zea} and EPOS-LHC~\cite{Pierog:2013ria}, to simultaneously reproduce all the observed behaviors for small systems. To understand the origins of these phenomena, event shape observables such as unweighted transverse spherocity ($S_{\rm 0}^{p_{\rm T}= 1}$) and the relative transverse activity classifier ($R_{\rm{T}}$) can be exploited as powerful tools to separate events dominated by hard and soft particles. Here, we report the production of light flavor hadrons for different classes of $S_{\rm 0}^{p_{\rm T} = 1}$ and $R_{\rm{T}}$ in pp collisions at $\sqrt{s}$ = 13 $\textrm{TeV}$. In addition, the system size dependence of charged particle production in pp, p--Pb and Pb--Pb collisions at  $\sqrt{s_{\rm NN}}$ = 5.02 TeV as a function of $R_{\rm{T}}$ is reported. Finally, we present a search for jet quenching behavior in small collision systems.

\section{Event shape observables}
Event shape observables such as $S_{\rm 0}^{p_{\rm T} = 1}$ and $R_{\rm T}$ have the capability to separate events with back-to-back jet structures from events dominated by multiple soft scatterings.

\subsection{Unweighted transverse spherocity ($S_{\rm 0}^{p_{\rm T} = 1}$)}
The unweighted transverse spherocity is given by
\begin{eqnarray}
S_{0}^{p_{\rm T} = 1} = \frac{\pi^{2}}{4} \min_{\hat{n}} \bigg(\frac{\Sigma_{i}~|p_{\rm T_{i}}\times\hat{n}|}{N_{\rm trks}}\bigg)^{2}.
\label{eq1}
\end{eqnarray}
Here, $\hat{n}$ is a unit vector that minimizes Eq.~\ref{eq1} and $N_{\rm trks}$ is the total number of charged-particle tracks in a given event. $S_{0}^{p_{\rm T} = 1}$ is calculated using charged-particle tracks that have $p_{\rm T} > $ 0.15 GeV/$c$ with at least 10 charged particle tracks in an event to  ensure  that  the  concept  of a topology is statistically meaningful. The charged-particle tracks are reconstructed using the Time Projection Chamber (TPC), within the pseudorapidity interval $|\eta|<0.8$. Unlike the estimator discussed in Ref.~\cite{Acharya:2019mzb}, the $p_{\rm T}$ of each track is normalized to 1 ($p_{\rm T}$ = 1) to minimize biases which affect neutral particle yields. By constuction, the $S_{0}^{p_{\rm T = 1}}$ estimator varies between the values 0 and 1. Here, the two extreme limits correspond to the two different topological limits. Events with $S_{0}^{p_{\rm T} = 1} \rightarrow$ 0 mostly consist of a single back-to-back jet while events with $S_{0}^{p_{\rm T} = 1} \rightarrow$ 1 are dominated by isotropic particle production. From here onwards, the events located in the bottom 20\% of the $S_{0}^{p_{\rm T} = 1}$ distribution are referred as jetty events while the top 20\% of the $S_{0}^{p_{\rm T} = 1}$ distribution are referred as isotropic events. As strangeness enhancement is observed in high-multiplicity pp collisions~\cite{ALICE:2017jyt}, a top 10\% high-multiplicity requirement is also imposed for the event selection (multiplicity class I-III~\cite{Acharya:2019mzb}). The multiplicity is estimated using the Inner Tracking System (ITS) at mid-rapidity ($|\eta|<0.8$) and is referred as the $N_{\rm SPD}$ estimator. Also, the multiplicity is estimated with the charged particle multiplicity at forward rapidity (2.8 $<\eta<$ 5.1 and -3.7 $<\eta<$ -1.7) using V0M scintillators, and is referred as the V0M estimator. One of the main differences in the measurements of particle production using two different estimators is that for the V0M multiplicity estimator, the pseudo-rapidity regions for multiplicity estimation and for the measurement of the particle spectra are different, while for the $N_{\rm SPD}$ multiplicity estimator, the multiplicity and particle spectra are measured in the same rapidity region. 

\subsection{Relative transverse activity classifier ($R_{\rm T}$)}
Using the relative transverse activity classifier, the final-state particle production can be studied as a function of varying underlying events. To ensure that at least one hard scattering took place in the event, analysed events are required to have a leading trigger particle above a certain $p_{\rm T}$. An event can be classified into three different azimuthal regions, relative to the trigger particle. Assuming $\phi_{\rm trig.}$ as the azimuthal angle for the leading trigger particle and $\phi_{\rm assoc.}$ as the azimuthal angle of the associated particles, the regions are classified as the following,
\begin{itemize}
\item Near/Toward-side: $|\phi_{\rm trig.} - \phi_{\rm assoc.}| < \frac{\pi}{3}$ 
\item Away-side: $|\phi_{\rm trig.} - \phi_{\rm assoc.}| > \frac{2\pi}{3}$
\item Transverse-side: $\frac{\pi}{3} \leq |\phi_{\rm trig.} - \phi_{\rm assoc.}| \leq \frac{2\pi}{3}$
\end{itemize}
Particle production in the near-side is dominated by jet fragmentation and the away-side region consists of some of the back-scattered jets. The transverse region is mostly dominated by particles produced by the underlying event (UE). Note that both the near- and away-side regions also contain similar UE production, which mean that one can subtract it, see Sec.~\ref{IAA}. The leading-$p_{\rm T}$ selection of  $>$ 5 GeV/$c$ ensures that the number density in the transverse region remains almost independent of leading particle $p_{\rm T}$~\cite{Acharya:2019nqn}. For the analysis on identified particle production the leading-$p_{\rm T}$ selection of $\geq$ 5 GeV/$c$ is considered, while for the study on the search of jet-quenching effects a leading-$p_{\rm T}$ selection of 8 $< p_{\rm T}^{\rm trig.} <$ 15 GeV/$c$ is considered, which reduces the sensitivity to elliptic flow. The relative transverse activity classifier ($R_{\rm T}$) is defined as~\cite{Martin:2016igp,Ortiz:2017jaz},
\begin{eqnarray}
R_{\rm T} = \frac{N_{\rm ch}^{\rm TS}}{\langle N_{\rm ch}^{\rm TS} \rangle}.
\label{eq2}
\end{eqnarray}
Here, $N_{\rm ch}^{\rm TS}$ is the charged particle multiplicity in the transverse region. The events with $R_{\rm T} \rightarrow$ 0 are the events with little or no UE and they are expected to be dominated by jet fragmentation.

\section{Results and Discussion}

\subsection{Identified particle production as a function of $S_{\rm 0}^{p_{\rm T}= 1}$ in pp collisions at $\sqrt{s}$ = 13 TeV}

\begin{figure}[h]
\centering
\includegraphics[width=12pc]{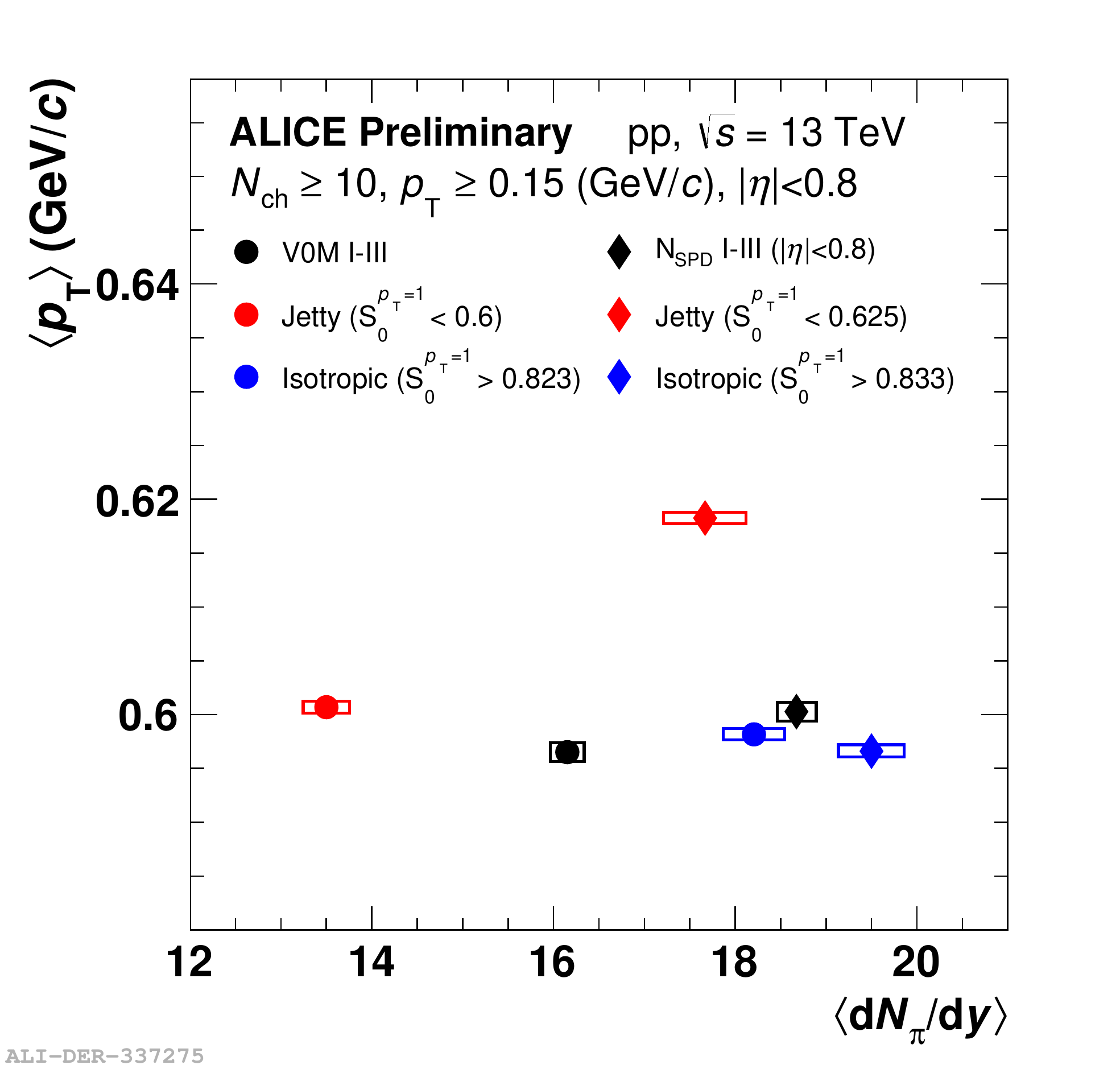}
\begin{minipage}[b]{14pc}\caption{\label{fig1} Pion $\langle p_{\rm T}\rangle$ vs integrated yield for different multiplicity estimators in different $S_{\rm 0}^{p_{\rm T = 1}}$ classes for pp collisions at $\sqrt{s}$ = 13 TeV.}
\end{minipage}
\end{figure}

\begin{figure}[h]
\centering
\includegraphics[width=12pc]{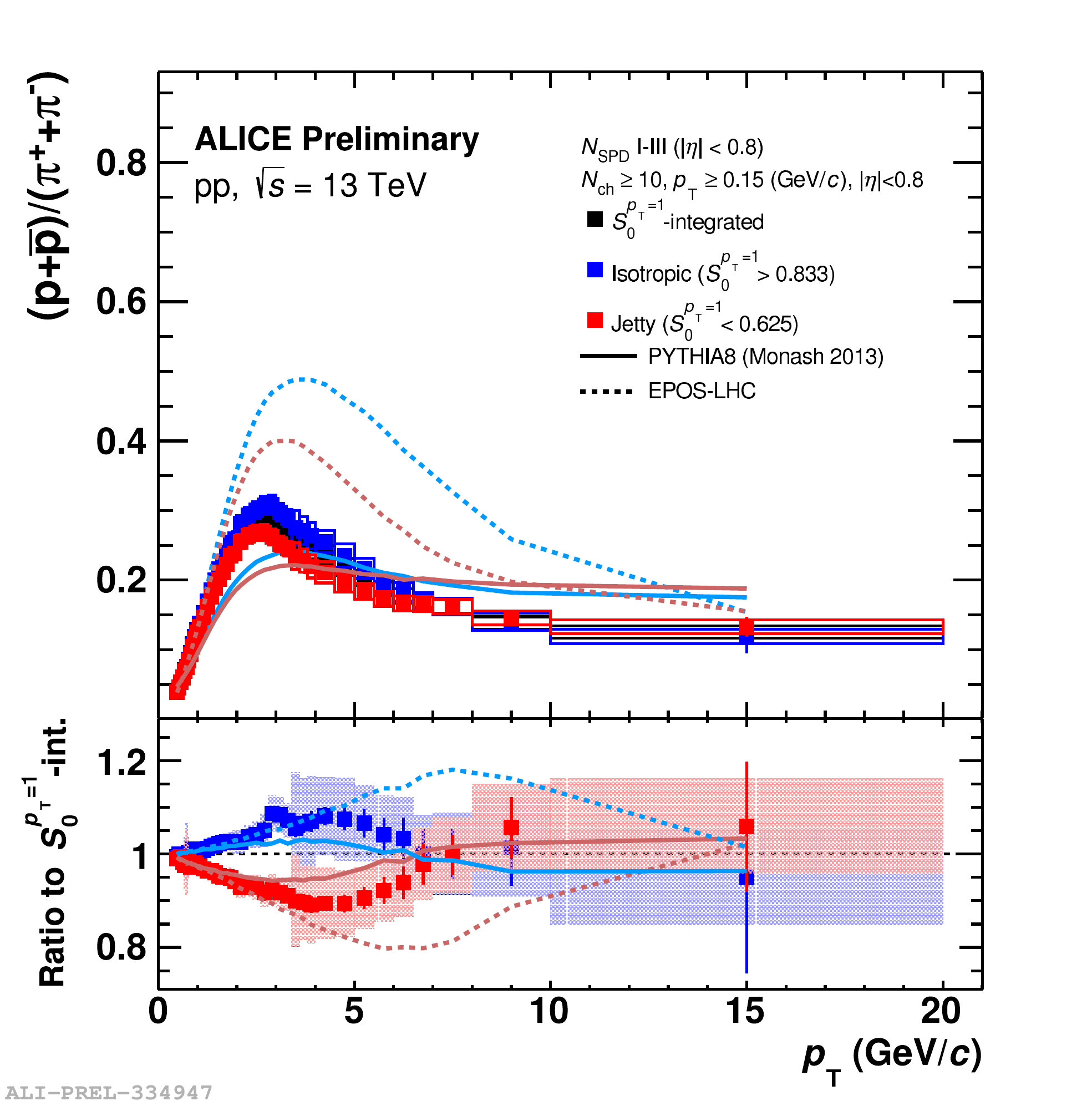}
\includegraphics[width=12pc]{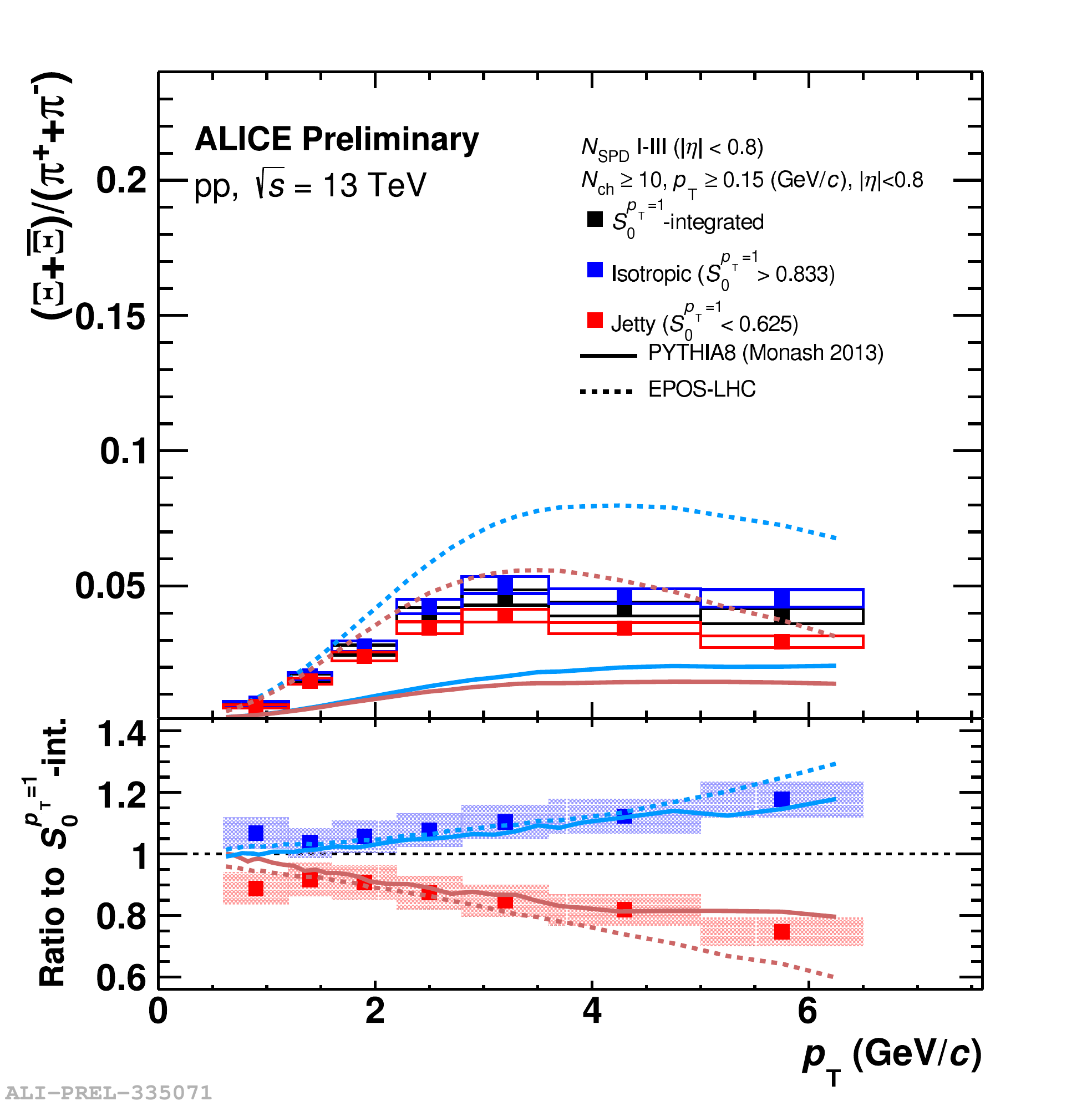}
\caption{\label{fig2} proton-to-pion (left) and $\Xi$-to-pion (right) ratios as a function $p_{\rm T}$ in different $S_{\rm 0}^{p_{\rm T}= 1}$ classes for pp collisions at $\sqrt{s}$ = 13 TeV, where the multiplicity selection is done using the $N_{\rm SPD}$ multiplicity estimator. The bottom panels show the double ratio of particle ratios from isotropic and jetty events to the $S_{\rm 0}^{p_{\rm T}= 1}$-integrated events. The ratios are compared with predictions from PYTHIA 8 and EPOS-LHC.}
\end{figure}
 Figure~\ref{fig1} shows the measured pion $\langle p_{\rm T}\rangle$ vs integrated yield for events classified using different multiplicity estimators, in different $S_{\rm 0}^{p_{\rm T}= 1}$ classes for pp collisions at $\sqrt{s}$ = 13 TeV. For the events selected based on the V0M estimator, isotropic and jetty events have a larger pion multiplicity difference, while the $N_{\rm SPD}$-triggered events disentangle soft and hard events more accurately (higher difference of $\langle p_{\rm T}\rangle$ of pions) in relatively smaller multiplicity gap of pions compared to the V0M estimator. Figure~\ref{fig2} shows the proton-to-pion (left) and $\Xi$-to-pion (right) $p_{\rm T}$-differential particle ratios in different $S_{\rm 0}^{p_{\rm T}= 1}$ classes, where the multiplicity selection was done using the $N_{\rm SPD}$ multiplicity estimator. The bottom panels show the double ratio of particle ratios from isotropic and jetty events to the $S_{\rm 0}^{p_{\rm T}= 1}$-integrated events. Here, pions and protons are identified using the specific energy loss ($\rm{d}E/\rm{d}x$) measured by the TPC and time-of-flight (TOF) using the TOF detector. $\Xi$ ($\rightarrow \pi + \Lambda, \Lambda \rightarrow \pi + p $) baryons are reconstructed from their topological decay properties and their yield is extracted from the invariant mass distribution of the decay products. The proton-to-pion ratio show larger enhancement in intermediate-$p_{\rm T}$ for isotropic events compared to jetty events, which seems to be reminiscent of flow effects in Pb-Pb collisions~\cite{Acharya:2018orn}. The $\Xi$-to-pion ratio suggests that the strange particle production with respect to non-strange particles is higher in isotropic events compared to jetty events. In general, the particle ratios are not described by PYTHIA8 and EPOS-LHC while the trends of double ratios are better described by both the event generators.
 
 \subsection{Identified particle production as a function of $R_{\rm T}$ in pp collisions at $\sqrt{s}$ = 13 TeV}
  \begin{figure}[h]
\centering
\includegraphics[width=12pc]{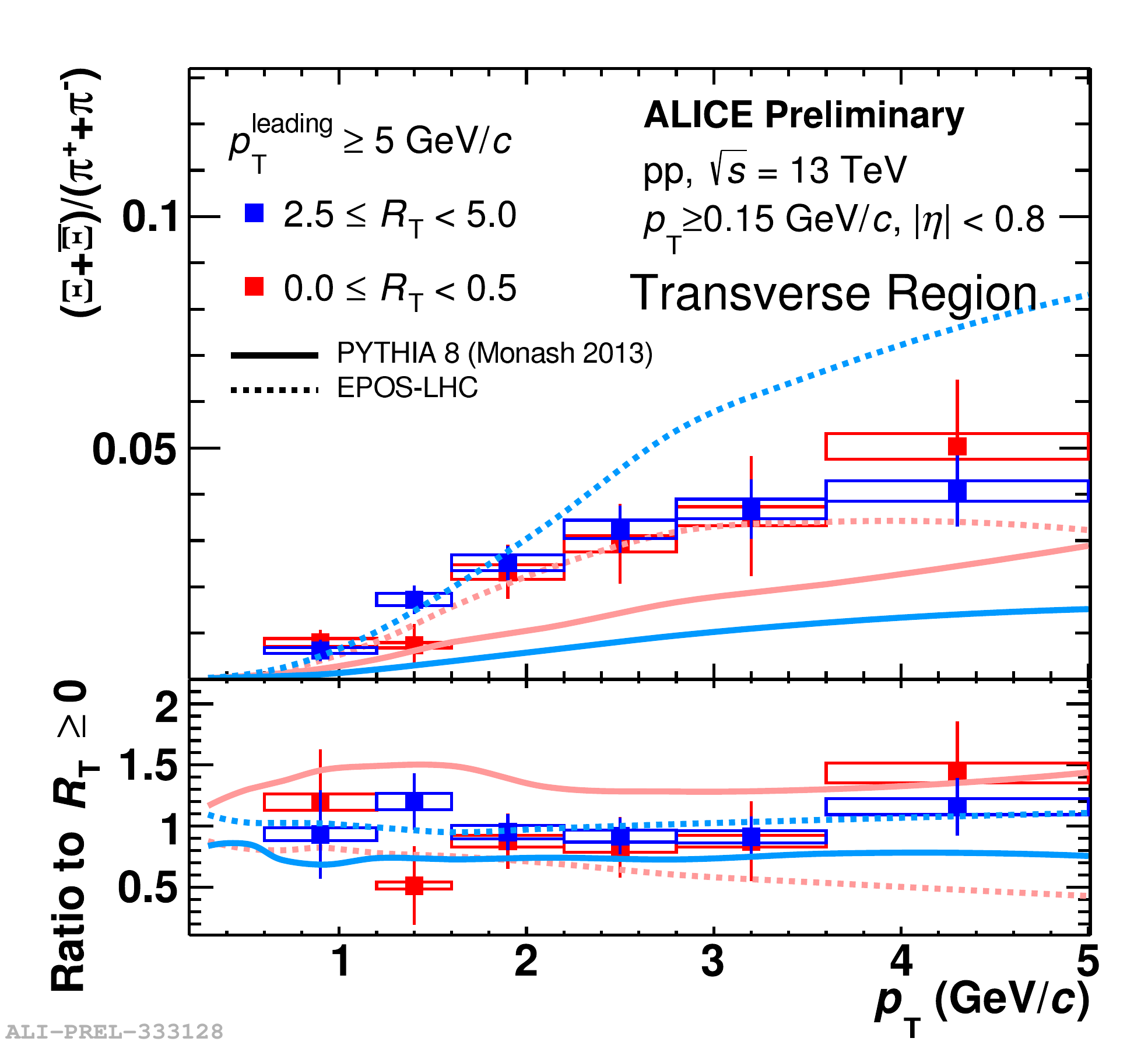}
\includegraphics[width=12pc]{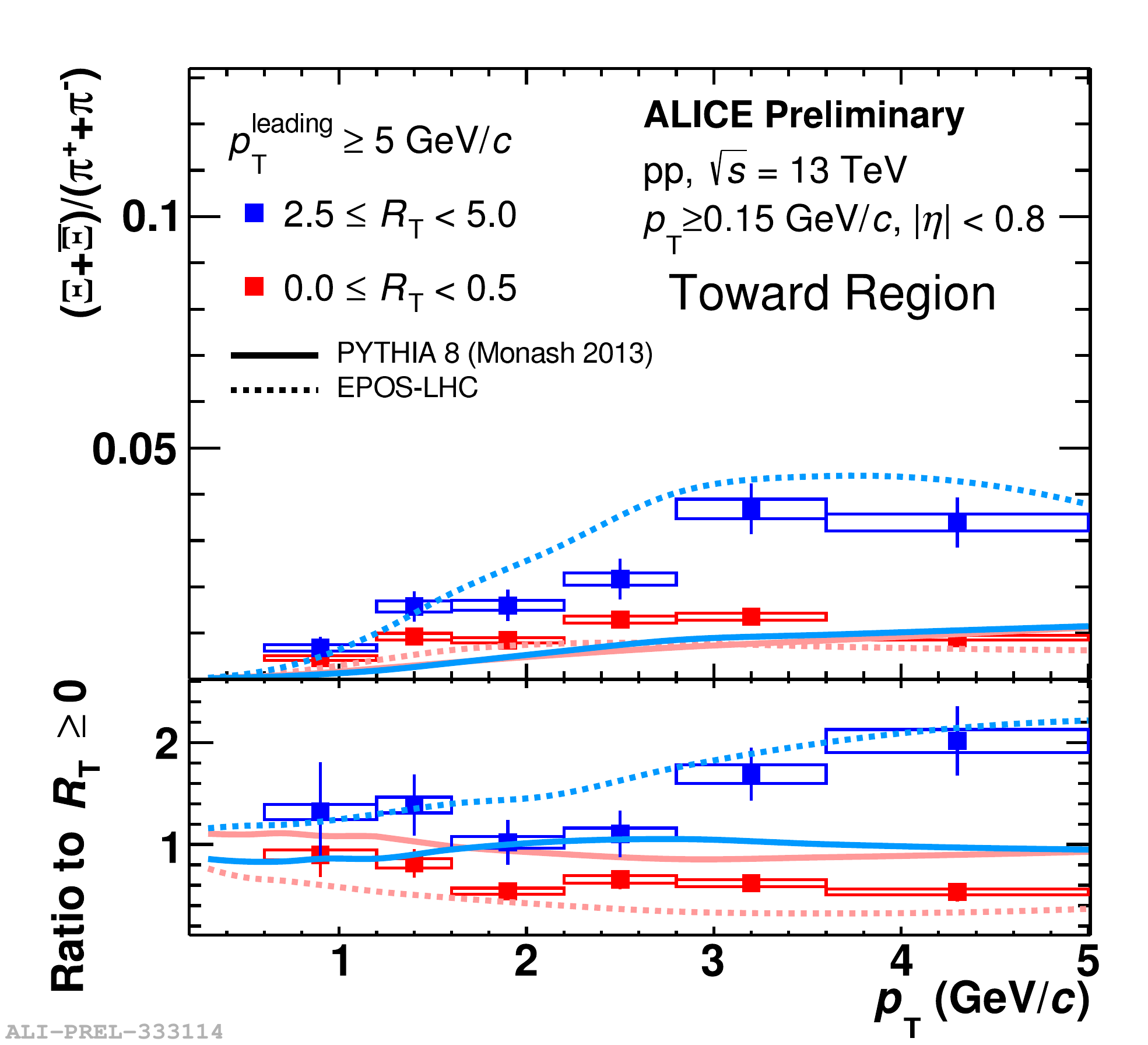}
\caption{\label{fig3} $\Xi$ to pion ratio as a function $p_{\rm T}$ in different $R_{\rm T}$ classes for the transverse (left) and toward (right) region for pp collisions at $\sqrt{s}$ = 13 TeV. The bottom panels show the double ratio of particle ratios from isotropic and jetty events to the $R_{\rm T}$-integrated events. The ratios are compared with predictions from PYTHIA8 and EPOS-LHC.}
\end{figure}
 Figure~\ref{fig3} shows the $\Xi$ to pion ratio as a function of $p_{\rm T}$ in different $R_{\rm T}$ classes for the transverse (left) and toward (right) region for pp collisions at $\sqrt{s}$ = 13 TeV. The bottom panels show the double ratio of particle ratios from isotropic and jetty events to the $R_{\rm T}$-integrated events. The ratio in the transverse region is nearly independent of $R_{\rm T}$ while the predictions from EPOS-LHC and PYTHIA8 show significant differences in two different $R_{\rm T}$ classes. However, for the toward region the ratio highly depends on $R_{\rm T}$ and EPOS-LHC seems to describe the data. This behavior suggests that the relative $\Xi$-baryon production is lower in events dominated by jet-fragmentation compared to those dominated by underlying events, indicating strange particle production is enhanced in the underlying event. This behavior supports the similar trend when studied in different $S_{\rm 0}^{p_{\rm T}= 1}$ classes.

\subsection{System size dependence of charge particle production as a function of $\langle N_{\rm ch}^{\rm TS} \rangle$} 
\label{IAA}
\begin{figure}[h]
\centering
\includegraphics[width=12pc]{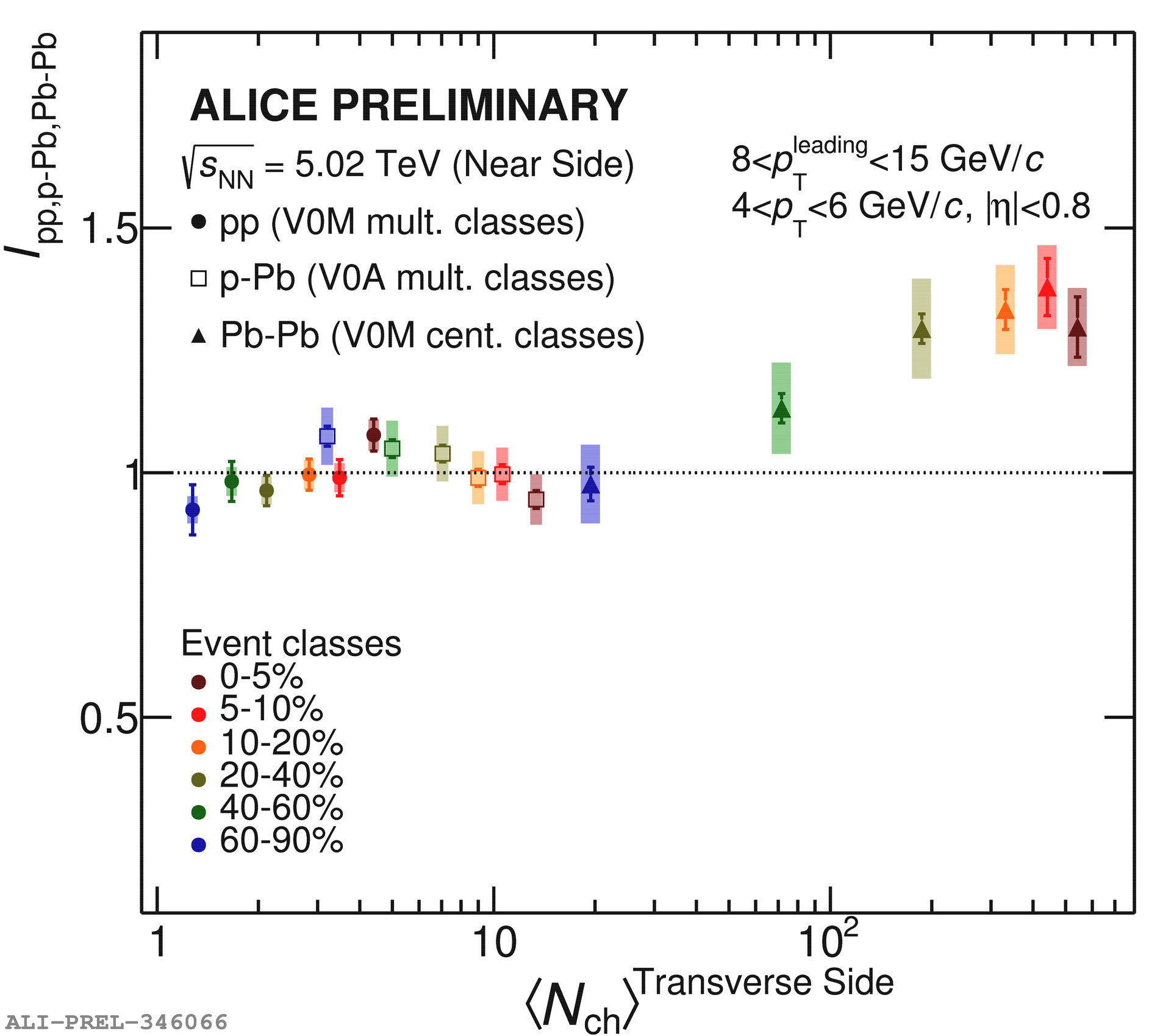}
\includegraphics[width=12pc]{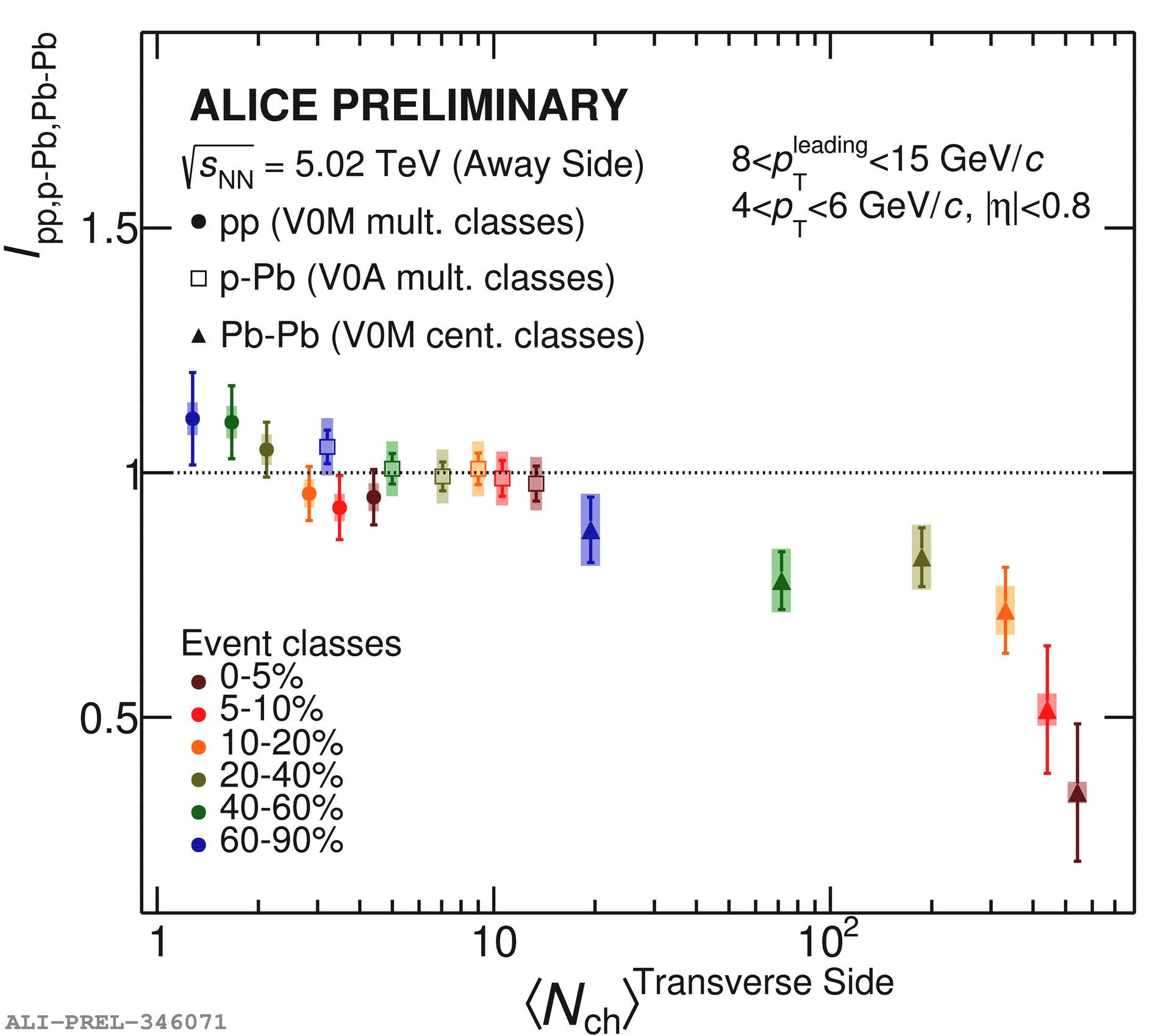}
\caption{\label{fig5} $I_{\rm pp,p-Pb,Pb-Pb}$ as a function of $\langle N_{\rm ch}^{\rm TS} \rangle$ in different V0M/V0A multiplicity classes for the near- (left) and away- (right) side in pp, p--Pb, and Pb--Pb collisions at $\sqrt{s_{\rm NN}}$ = 5.02 TeV.}
\end{figure}
To explore the presence of jet-quenching effects we have calculated the $I_{\rm pp,p-Pb,Pb-Pb}$, an observable which is calculated from the yields of different topological regions, as a function of $\langle N_{\rm ch}^{\rm TS} \rangle$ for different V0M/V0A multiplicity classes of pp, p--Pb and Pb--Pb collisions. The $I_{\rm pp,p-Pb,Pb-Pb}$ is an analogous quantity calculated as in Ref.~\cite{Aamodt:2011vg}, which is sensitive to medium effects. The suppression of this observable in the away side would indicate the presence of jet quenching, while an enhancement in the near side would indicate the presence of medium effects and bias due to trigger particle selection.  It is defined as the ratio of the yield in the near or away region (after subtraction of underlying events) in different collision systems to the yield in near or away region in minimum bias pp collisions. As a direct selection on $N_{\rm ch}^{\rm TS}$ biases the near- and away-side yields~\cite{Ortiz:2020dph}, here the events are selected based on a forward-rapidity estimator (V0M for pp and Pb--Pb collisions and V0A for p--Pb collisions) and the corresponding $N_{\rm ch}^{\rm TS}$ are calculated. Figure~\ref{fig5} shows the $I_{\rm pp,p-Pb,Pb-Pb}$ for the $p_{\rm T}^{\rm assoc.}$ range 4 $< p_{\rm T}^{\rm assoc.} < $ 6 GeV/$c$ as a function of $\langle N_{\rm ch}^{\rm TS} \rangle$ in different V0M/V0A multiplicity classes for the near (left) and away (right) side in pp, p--Pb, and Pb--Pb collisions at $\sqrt{s_{\rm NN}}$ = 5.02 TeV. In contrast to Pb-Pb collisions, for small collision systems, no enhancement (suppression) of $I_{\rm pp,p-Pb}$ is observed in the near (away) sides for pp, p--Pb collisions. This may indicate of the absence of jet-quenching effects in small collision systems for the measured  $\langle N_{\rm ch}^{\rm TS} \rangle$ ranges.

\section{Summary}
This work suggests that by using event shape observables like $S_{\rm 0}^{p_{\rm T} = 1}$ and $R_{\rm T}$, one can vary the magnitude of the underlying events and study the events separately with different topological limits (jetty vs isotropic). A clear dependence of light flavor particle $p_{\rm T}$ spectra and ratios on $S_{\rm 0}^{p_{\rm T}= 1}$ and $R_{\rm T}$ is observed. These studies indicate that the the strange particle production with respect to non-strange particles is higher in the events dominated by underlying events compared to the events dominated by jet fragmentation. In contrast to Pb--Pb collisions, no suppression of $I_{\rm pp,p-Pb}$ is observed in the away side for pp and p--Pb collisions, which indicates the absence of jet-quenching effects for small collision systems in the measured  $\langle N_{\rm ch}^{\rm TS} \rangle$ ranges.

\section*{Acknowledgments}
S.T. acknowledges the support from CONACyT under the Grant No. A1-S-22917 and postdoctoral fellowship of DGAPA UNAM.


\begin{thebibliography}{99}
\bibitem{ALICE:2017jyt} 
  J.~Adam {\it et al.} [ALICE Collaboration],
  Nature Phys.\  {\bf 13}, 535 (2017).
  
  \bibitem{Nagle:2018nvi}
J.~L.~Nagle and W.~A.~Zajc,
Ann. Rev. Nucl. Part. Sci. \textbf{68}, 211 (2018).

\bibitem{Sjostrand:2014zea}
T.~Sj\"ostrand {\it et. al.},
Comput. Phys. Commun. \textbf{191} (2015), 159.

\bibitem{Pierog:2013ria}
T.~Pierog,  {\it et. al.},
Phys. Rev. C \textbf{92}  034906 (2015).

\bibitem{Acharya:2019mzb}
S.~Acharya \textit{et al.} [ALICE Collaboration],
Eur. Phys. J. C \textbf{79} (2019) 857.

\bibitem{Acharya:2019nqn}
S.~Acharya \textit{et al.} [ALICE Collaboration],
JHEP \textbf{04} (2020), 192.

\bibitem{Martin:2016igp}
T.~Martin, P.~Skands and S.~Farrington,
Eur. Phys. J. C \textbf{76} (2016) 299.

\bibitem{Ortiz:2017jaz}
A.~Ortiz and L.~Valencia Palomo,
Phys. Rev. D \textbf{96} 114019 (2017).

  \bibitem{Acharya:2018orn} 
  S.~Acharya {\it et al.} [ALICE Collaboration],
  Phys.\ Rev.\ C {\bf 99}, 024906 (2019).

\bibitem{Aamodt:2011vg}
K.~Aamodt \textit{et al.} [ALICE Collaboration],
Phys.\ Rev.\ Lett.\  \textbf{108}, 092301 (2012).

\bibitem{Ortiz:2020dph}
A.~Ortiz, S.~Tripathy and G.~Benc\'edi,
[arXiv:2007.03857 [hep-ph]].

\end{thebibliography}
\end{document}